\documentclass[smallextended]{svjour3} 
\usepackage{amsmath}
\usepackage{graphicx}
\usepackage{color}
\usepackage[left]{lineno}

\RequirePackage{fix-cm}

\journalname{Transport in Porous Media}
\begin{document}

\title{Identifying space-dependent
coefficients and the order of fractionality in  fractional advection diffusion equation}

\titlerunning{Identifying 
parameters of fractional advection diffusion equation}
\author{Boris Maryshev             \and
Alain Cartalade    \and
Christelle Latrille       \and Marie-Christine N\'eel 
}

\authorrunning{B. Maryshev, A. Cartalade, C. Latrille and M.C. N\'eel}

\institute{ B. Maryshev\at Institute of Continuous Media Mechanics, UB RAS, Perm 614013, Russia, \email{bmaryshev@mail.ru}\\
%
           \and A. Cartalade \at Den-DM2S, STMF, LMSF, CEA, Universit\'e de Paris-Saclay, F-91191 Gif-sur-Yvette, France\\ \email{alain.cartalade@cea.fr}\\
\and C. Latrille \at Den-DPC, SECR, L3MR,  CEA,  Universit\'e de Paris-Saclay, F-91191  Gif-sur-Yvette, France\\ \email{Christelle.LATRILLE@cea.fr}
      \and     M.C. N\'eel \at
       EMMAH, INRA, Universit\'e d'Avignon et des Pays de Vaucluse, 84000, Avignon, 
   France\\  \email{mcneel@avignon.inra.fr}  
}

\date{Received: date / Accepted: date}
\maketitle

\begin{abstract}
  Tracer tests  in  natural porous media sometimes show abnormalities that suggest considering  a  fractional  variant of the Advection Diffusion Equation supplemented by a  time derivative
 of non-integer order. 
 We are describing an inverse method  for this equation: it finds  the order of the fractional derivative and the coefficients  that  achieve minimum discrepancy between solution and tracer  data. Using an adjoint equation  divides the computational
 effort by an amount proportional to the number of freedom degrees, which becomes large 
 when some coefficients  depend on space. 
Method accuracy is checked on 
 synthetical data, and applicability to actual tracer test is
 demonstrated.
\end{abstract}


\section{Introduction}

\label{intro}

In many natural media (river flows, aquifers, soils, porous columns) solute decay seems adequately described by models accounting for immobile fluid 
fraction \cite{Coats,Deans,Baker,Genuch,Schumer,Benson}. Such models
 are equivalent to  classical Advection-Dispersion Equation equipped of   supplementary operator compounding  time derivative and convolution. Exponential or algebraic convolution
 kernels yield  classical or fractional
 Mobile/Immobile Model \cite{Coats,Deans,Baker,Genuch} \cite{Schumer,Benson}. Though both variants  describe experimental
 break-through curves  \cite{Schumer,Gaudet} showing non-symmetric ascending and descending slopes,  they exhibit dramatically different late time behaviors. Hence,
predicting the future of a contamination event requires accurate parameter identification for each candidate model. In view of such prediction, we concentrate our attention on a method 
adapted to data recorded during limited time (differently from \cite{Tuan}) and  on the fractional MIM
\begin{linenomath*}
\begin{equation}
-\partial_{x}\left(\partial_{x}p_1u-Vu\right) +\partial_{t}p_2u+\partial_{t}^\alpha p_3 u={\mathcal R}(x,t).
  \label{finalEq}
\end{equation}
\end{linenomath*}
In this equation   $\partial_{t}^\alpha \equiv \partial_{t}I_{0,+}^{1-\alpha} $ is a derivative of  order $\alpha$ related to  the temporal convolution 
$ I_{0,+}^{1-\alpha}f(t)=\frac{1}{\Gamma\left(1-\alpha\right)}
\int_{0}^{t}\frac{f\left(t'\right)}{\left(t-t'\right)^{\alpha}}dt'$, itself called a fractional integral  \cite{Samko}, where $\Gamma$
is   the Euler gamma function, $\alpha$  belongs to $]0,1[$ and ${\mathcal R}$ is a source term.
This fractional generalization  \cite{Schumer,Benson} of  the Advection-Dispersion Equation  describes mass transport in one-dimensional media (e.g. rivers  \cite{Schumer,Haggerty} or the 
flow geometry considered by \cite{Young}) where  fluids 
can be temporarily immobile and  retain solutes during random trapping times of infinite average: in a porous medium, $p_1$, $p_2$ and $V$ represent  a dispersion coefficient,  the mobile volume fluid fraction and the water flux
density (or Darcy velocity). Coefficient $p_3$, proportional to the immobile fluid fraction,  is discussed in Section \ref{mod}. 

\begin{figure}
\scalebox{0.4}{\includegraphics*{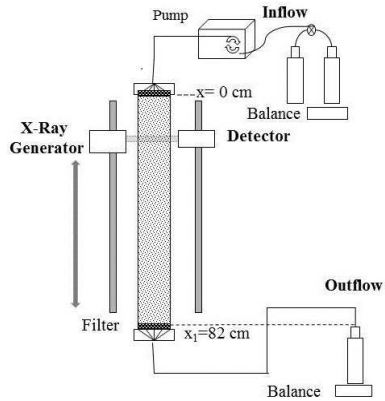}}
\caption{\label{exp}Di-chromatic X-ray spectrometry (DXS) device measuring water content and solute concentration at several cross-sections of a  column filled of unsaturated sand. The environment 
of the column (mottled)  achieves constant Darcy velocity $V$ and  water content
$\theta$, and injects  tracer flux $VC_0$ ($V=1.05$ cm/h, $C_{0}=0.10$ mol/l)  at  inlet $x=0$ between time instants $0$
 and $t_f=3$h. X ray generator and detector moved along the column measure tracer concentration and water content averaged on each desired cross-section. }
\end{figure}

\begin{figure}
\scalebox{0.45}{\includegraphics*{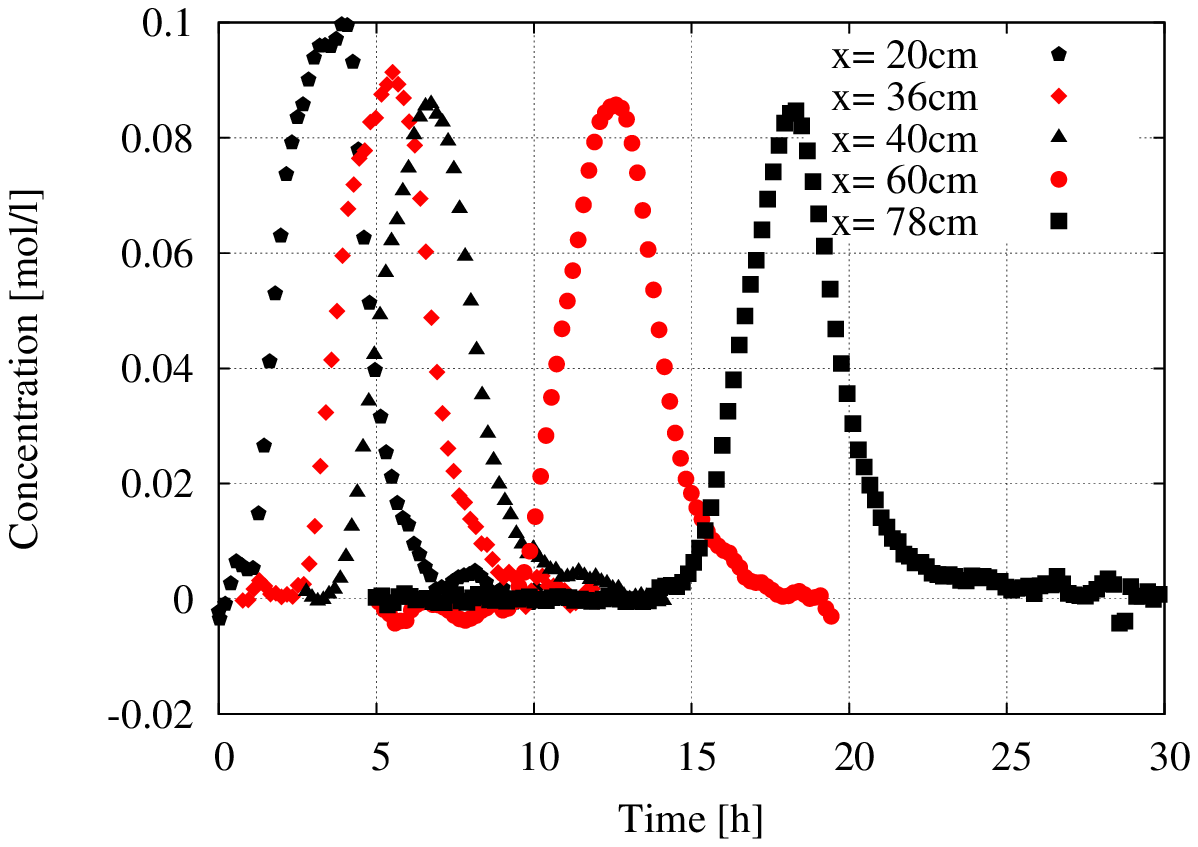}}\scalebox{0.45}{\includegraphics*{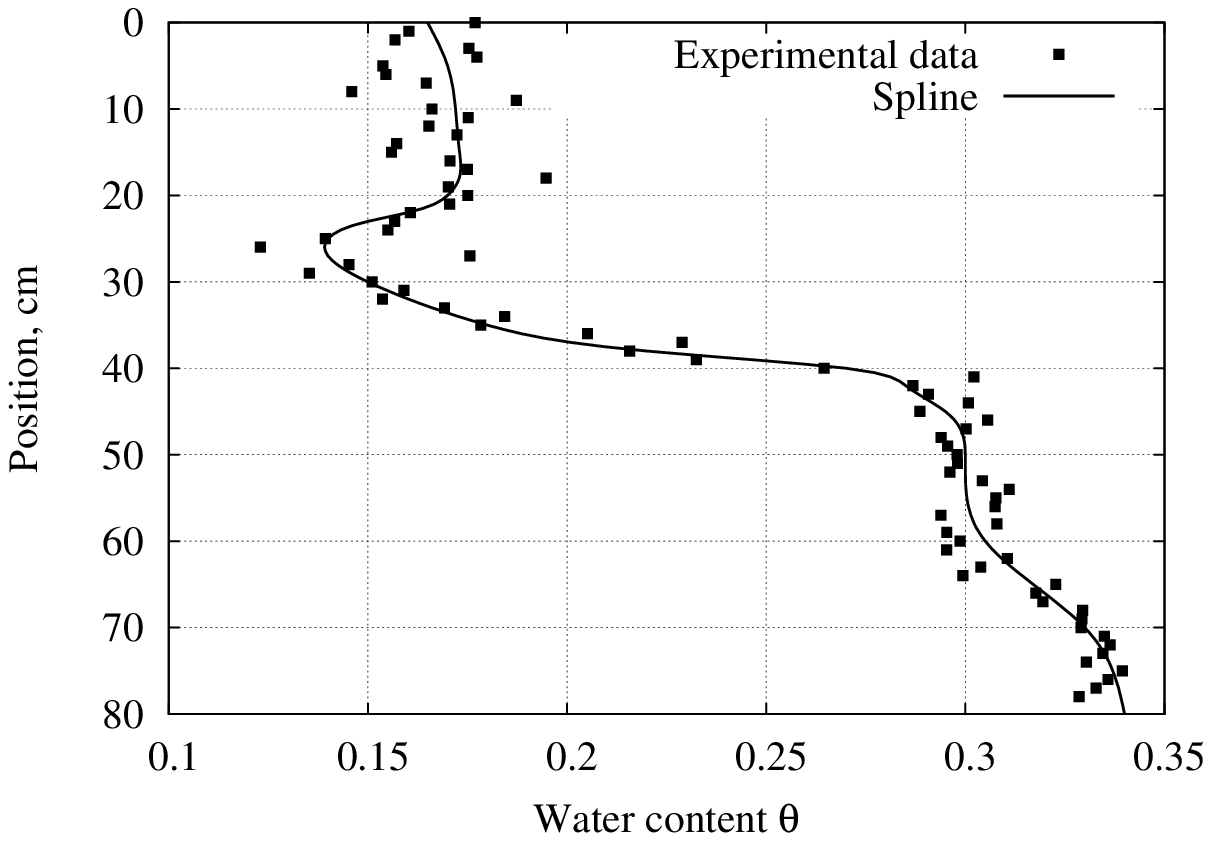}}
\caption{\label{WC_exp}Solute concentration and steady  water content measured in the device represented on Fig. \ref{exp}. Left:
 solute concentration versus time, measured at some of the $15$ inspected cross sections of the column, and represented by different
 symbols.  Right:
 water content $\theta$  versus
position: dots and full line represent measured  $\theta$ and spline   interpolation with $n_\theta=18$ knots, less spaced in the  poorly saturated region where $\theta$ exhibits stronger variations.
Interpolation issued from this   $n_\theta$ value coincides with a local average of $\theta$. }
\end{figure}

We ultimately aim to check the validity of Eq. (\ref{finalEq}) for solute transport in unsaturated sand on the basis of
 concentration profiles 
recorded at several cross-sections 
 of a column (of length $L$)
filled of such a medium \cite{Latrille,Latrille12} (see Fig.\ref{exp}).  Some of these profiles are  displayed on the left  of Fig.\ref{WC_exp}. However, transport 
properties are often sensitive to  water content $\theta$
 (see \cite{Gaudet,Padilla}),  measured   and  found steady:  the right panel of Fig.\ref{WC_exp} shows that $\theta$  depends on space. 
 Since  $\theta$ may influence some coefficients of (\ref{finalEq}),
 we account for possible variations by  linear interpolation using
 $n+1$  nodes including both ends of interval $[0,L]$. We do not know the most appropriate value of $n$, and  try 
several  interpolation sequences. In each  attempt
we estimate the best fit between data and  Eq.(\ref{finalEq})  equipped of piecewise continuous coefficients. The latter are determined by their interpolation
    values which play the role of  supplementary parameters to estimate: 
 the right panel of Fig.\ref{WC_exp} suggests a number of nodes  resulting in at least  forty  effective parameters.
 We store $\alpha$, the uniform coefficients (if there are) and the interpolation values of those which  depend on space
 in parameter vector $\textbf{q}$ which determines a  solution  of  the discrete direct problem, namely a discrete version of Eq.(\ref{finalEq}). The smallest possible squared distance
  $E(\textbf{q})$ between this solution and the data  gives the best candidate  for $\textbf{q}$ and quantifies the discrepancy between model and experiment.

\begin{figure}
\begin{center}\includegraphics[width=0.6\linewidth]{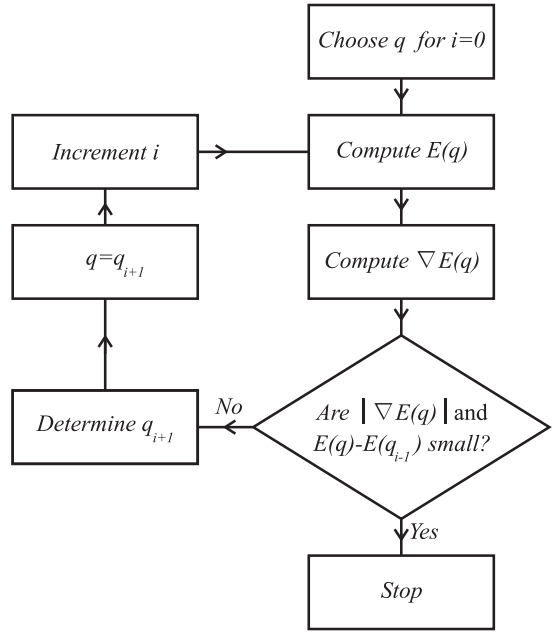}
\end{center}
\caption{\label{flowchart} Principle of an optimization process  minimizing $E(\textbf{q})$ by building   a sequence of parameters ($\textbf{q}_i$). Each
 next element $\textbf{q}_{i+1}$ of the sequence is
 determined by action ``Determine  $\textbf{q}_{i+1}$'' on the basis of  inputs provided by actions ``Compute $ E(\textbf{q})$'' and ``Compute $\nabla E(\textbf{q})$'' applied to $\textbf{q}_{i}$.}
\end{figure}

 We find  this minimum  at the end of a  sequence ($\textbf{q}_i$)  in parameter space: each $\textbf{q}_{i+1}$ is the issue of
task  `` Determine $\textbf{q}_{i+1}$'' in the optimization loop schematically represented on  Fig.\ref{flowchart}. A robust and accurate
algorithm \cite{Nocedal} completes this task by deducing $\textbf{q}_{i+1}$ from $\textbf{q}_i$ and from the  $E(\textbf{q}_j)$ and  ${\nabla}E(\textbf{q}_j)$ issued
 from last and penultimate steps.
Hence, action
 ``Compute ${\nabla}E(\textbf{q})$'' needs to be  accurate and quick. Finite differences successively incrementing each entry of vector
  $\textbf{q}_i$  would 
 necessitate solving at least  $2n+4$ copies of the discretized p.d.e. (\ref{finalEq})  for each repetition of this task. Hence, we prefer   the adjoint state method \cite{Chavent,Chavent5,Sun}
that instead  solves one adjoint equation exactly as complex as the direct discrete problem: this  divides by  at least $n+2$ the computing time  necessary for each 
  repetition of the loop. 

The fractional model (\ref{finalEq}) and  the optimization problem  are detailed in Section \ref{Sec2}. 
Section \ref{AdjStM} defines the adjoint equation that gives us the
gradient of  $E(\textbf{q})$. 
This sets the principle of an inversion method. 
 A numerical experiment applied on synthetic data confirms that it  accurately retrieves the   coefficients of (\ref{finalEq}).  Section \ref{App} demonstrates that this
 approach applies  to actual  tracer test.

\section{Mathematical model (\ref{finalEq}) and  optimization problem associated with data}\label{Sec2}
The time concentration profiles represented  on Fig.\ref{WC_exp} were recorded in a partly saturated sand column in which 
  the water content $\theta$ was found constant.
 While the Darcy velocity  $V$  was also measured,  $p_1$, $\alpha$, $p_2$ and $p_3$ could not be measured. Before  estimating them  by minimizing the discrepancy between
 the records
and a numerical solution of  (\ref{finalEq}) associated to boundary conditions representing the experiment, we discuss the links between measured quantities, model and parameters.

\subsection{Model}
\label{mod}

 The several versions of the Mobile-Immobile Model\cite{Coats,Deans,Baker,Genuch} assume two fluid states (mobile and immobile) occupying volume fractions $p_2$ and $\theta_{im}$ 
of the medium. Solute concentrations are $u$ and  $C_{im}$ in these two states. In water flowing through porous media, the total fluid fraction  $p_2+\theta_{im}$ is the water
 content $\theta$.  Dichromatic X ray Spectroscopy measures this quantity and  the total solute concentration $C$ \cite{Latrille,Latrille12} which satisfies
$\theta C=p_2u+\theta_{im}C_{im}$. Yet,  $u$, $C_{im}$, $p_2$ 
and $\theta_{im}$ are not measured.

The original version of the MIM assumes Fick's law in mobile phase and   exchanges with  immobile phase obeying first order kinetic  equivalent to taking
$\partial_tC_{im}$ proportional to $ u-C_{im}$. At molecular level this model   is  equivalent to Brownian motion interrupted during  exponentially
distributed time lapses \cite{Valo89} and its
 solutions  decay  exponentially at late times. Hence, a look at the 
tailings exhibited by the concentration records displayed on Fig.\ref{WC_exp}  suggests examining  the fractional variant (\ref{finalEq}) that exhibits algebraic asymptotic behavior.
Nevertheless, we do not use these tailings to discriminate between the two variants because they exhibit negative records  revealing large relative errors.
  The fractional variant is equivalent to
 release times distributed
 by stable subordinator
of stability exponent $\alpha$ in $]0,1[$, an assumption that implies \cite{Schumer,Benson}
\begin{linenomath*}
\begin{equation}
   C=\frac{1}{\theta}\left(p_2{\mathrm{Id}}+p_3I_{0,+}^{1-\alpha}\right)u\equiv b(\theta, p_1,p_2,p_3,\alpha,u),
  \label{totC}
\end{equation}
\end{linenomath*}
 where  $p_3=\Lambda p_2$ also depends on 
 $\Lambda=\lambda {\mathcal H}$.  The latter quantity incorporates a scale factor $\lambda$ of the dimensionality of  $[T]^{1-1/\alpha}$ and the probability ${\mathcal H}$ of being immobilized \cite{NeelZ}. 
 Fick's law applied to the mobile concentration  yields solute flux equal to
\begin{linenomath*}
\begin{equation}
vp_2u-\partial _{x}(p_1u),
 \label{Fick}
\end{equation}
\end{linenomath*}
where  $v$ represents  a local average velocity of  particles in  mobile state.  
 Inside porous columns
$vp_2$ is commonly assumed to be equal to the Darcy velocity $V$  \cite{Sardin}, a measured 
quantity which does not depend on $x$.  Eqs.(\ref{totC}) and  (\ref{Fick}) imply (\ref{finalEq}) with ${\mathcal R}=0$. 
 
\subsection{Boundary conditions}
\label{bou}
Tracer solution of concentration $C_0$ injected with the fluid at  flow rate $V$ at  column  inlet $x=0$    between time instants
 $t=0$ and $t_{f}$ results into  tracer flux  rate $ VC_{0}H(t)H(t_{f}-t)$ at $x=0$, $H$ representing the Heaviside function. We  assume zero diffusive flux at the outlet
 as \cite{Neupauer99,SanduG}, and homogeneous  initial condition meaning that the system is initially free of tracer:
\begin{linenomath*}
\begin{equation}
  Vu-{\partial _x}(p_1u) |_{x=0}=VC_{0}H(t)H(t_{f}-t),\quad {\partial _x}(p_1u)|_{x=L} =0, \quad u|_{t=0}=0.
\label{contibc}
 \end{equation}
\end{linenomath*}

\subsection{Degrees of freedom}
\label{dof}
Water content certainly influences the mobile water content, and the right panel of Fig.\ref{WC_exp} strongly suggests that $p_2$ and $p_3$ depend on $x$.
  Though the inverse method  presented here still works when $p_1$ also depends on $x$, we  consider this parameter
 uniform for sake of simplicity and discuss this choice at the end of Section \ref{sub:Calibration-of-experimental}.
  We approximate the unknown  functions $p_2(x)$  and $p_3(x)$   by linear interpolation based on
$n+1$  nodes 
 $x^{(0)}=0<x^{(1)}<...<x^{(n)}=L$ which we fix before starting parameter identification, as mentioned in the Introduction. In each  interval $[x^{(i)},x^{(i+1)}]$ 
with $0\leq i\leq n-1$ we impose linear variations
\begin{linenomath*}
\begin{equation}
  {p_j}( x)={p_j}^{(i)}+
\frac{{p_j}^{(i+1)}-{p_j}^{(i)}}{x^{(i+1)}-x^{(i)}}
\left (  x-x^{(i)} \right )\:\mbox{for}\: x^{(i)}\leq  x\leq x^{(i+1)}: 
  \label{interpb}
\end{equation} 
\end{linenomath*}$p_j( x)$ depends on ${p_j}^{(i)}$ only if $ x$ belongs to $[x^{(i-1)},x^{(i+1)}]$.
 Due to  (\ref{interpb}),  $p_1$, $\alpha$ and  the set of all the $p_j^{(i)}$ with $j=2,3$ and $i=0,...,n$ determine the solutions of problem (\ref{finalEq}-\ref{contibc}). Therefore, 
   we  store in vector \break $\textbf{q}\equiv (p_1,p_2^{(0)},...,p_2^{(n)},p_3^{(0)},...,p_3^{(n)},\alpha)^\dag$ the $2n+4$ effective parameters 
$p_1$, ${p_2}(x^{(0)})$,..., ${p_2}(x^{(n)})$, ${p_3}(x^{(0)})$,..., ${p_3}(x^{(n)})$, $\alpha$ which we rename $q_1$, $q_2$,..., $q_{2n+4}$.

\subsection{Discrete direct problem}
In fact, only numerical approximations to the solution $u$ of problem (\ref{finalEq}-\ref{contibc}) and to $ b(\theta,p_2,p_3,\alpha,  u)$ are available. Taking  $T>t_f$ we specify in Appendices 
 \ref{Discrint}-\ref{spe} the  approximations which we use  in the space-time domain $[0,\, L]\times[0,\, T]$ discretized by
 $N_{Sp}+2$ space nodes  $s\Delta x$ (including both ends of $[0,L]$) and $N_{T}+1$ time nodes $k\Delta t$ satisfying  $L=(N_{Sp}+1)\Delta x$ and  $T=N_T\Delta t$. 
The    $u_{s}^{k}$ that approximate  the $u\left(s\Delta x,\, k\Delta t\right)$
 for $s=1,...,N_{Sp}$ and 
$k=0,...,N_T$ constitute an array 
noted  $\textbf{u}$, of columns  $\textbf{u}^0,...,\textbf{u}^{N_T}$. We call $X$ the set of all arrays of $N_{Sp}$ lines and $N_T+1$ columns. 
Appendix \ref{spe} specifies a  linear algebraic problem  (the discrete problem) 
\begin{linenomath*}
\begin{equation}
  A(\textbf{q},\textbf{u})=\textbf{r}(\textbf{q}), \quad\textbf{u}\in {\mathcal D}( A)
  \label{Dis_sh}
\end{equation} 
\end{linenomath*}
that determines an approximation to the solution of (\ref{finalEq}-\ref{contibc}). We call $\textbf{u}_{\textbf{q}}$ the unique array of  ${\mathcal D}( A)\equiv\left\{\mathbf{u}\in X/\mathbf{u}^{0}=0\right\}$ solving  (\ref{Dis_sh})
for any specified  $\textbf{q}$ in the convex closed set  $ {Q}_\varepsilon= {R}_+\times[\frac{V\Delta t}{\Delta x}+\varepsilon,+\infty[^{n+1}\times  {R}_+^{n+1}\times[0,1]$.  
Appendix \ref{Diag_dom} shows that  $\textbf{u}_{\textbf{q}}$
 depends  smoothly on  $\textbf{q}$ when the latter belongs to  $ {Q}_\varepsilon$.

Inside porous media, instead of $u$ we measure the total solute concentration  and compare it with an approximation of $b(\theta, p_1,p_2,p_3,\alpha,u)$ defined in Eq. (\ref{totC}).
\subsection{Comparing model and  data} \label{comp}
For each $(s\Delta x,k\Delta t)$ in $]0,L[\times[0,T]$, 
the approximation 
\begin{linenomath*}
\begin{equation}
  \left(B(\textbf{q},\textbf{u})\right)_{s}^{k}\equiv
\frac{p_2}{\theta}(s\Delta x)u_{s}^{k}+\frac{p_3}{\theta}(s\Delta x)
\sum_{j=0}^k{\mathcal I}^{j, k}u_{s}^{k-j}
\label{totCdis}
\end{equation}
\end{linenomath*}
to $b(\theta, p_1,p_2,p_3,\alpha,u)(s\Delta x,k\Delta t)$ is consistent with that detailed in Appendix \ref{Discrint} for $I_{0,+}^{1-\alpha}$, and we store the $\left(B(\textbf{q},\textbf{u})\right)_{s}^{k}$  in array
  $\textbf{B}(\textbf{q}, \textbf{u})$. For  $j=2,3$ this expression involves the ${p_j}(s\Delta x)$  which we deduce from the 
entries of $\textbf{q}$  according to (\ref{interpb}). The ${\mathcal I}^{j, k}$ are defined in Appendix \ref{Discrint}, and the $\theta(s\Delta x)$ are given by the spline interpolation represented at
 the right of Fig.\ref{WC_exp}.

The total concentration $C$ is  measured on 
${\mathcal N}^{(d)}$ elements 
  $(\bar{s}\Delta x,\,\bar{k}\Delta t)$ of the discretization grid: their 
 indexes  $(\bar{s},\bar{k})$ form the subset ${\mathcal M}^{(d)}$ of $\left\{ 1,..., N_{Sp}\right\} \times \left\{ 1,..., N_{T}\right\}$. 
We furthermore impose $\bar{k}<N_T$, and  
 call $C_{\bar{s}}^{\bar{k}}$ the concentration recorded at position $\bar{s}\Delta x$ 
and time $\bar{k}\Delta t$. We store these records $C_{\bar{s}}^{\bar{k}}$ 
in an array $\textbf{C}$  whose each  entry of index
   not belonging to ${\mathcal M}^{(d)}$ is  set equal to zero. We compare $\textbf{C}$
 with  $\textbf{B}(\textbf{q}, \textbf{u})$  by
 normalizing with the injected solute concentration $C_0$
 distances issued from the standard Euclidean scalar 
product of $X$ (namely $\langle\textbf{u}\cdot\textbf{w}\rangle_{X}=$ $\sum_{s=1}^{N_{Sp}}\sum_{k=0}^{N_{T}}u_{s}^{k}w_{s}^{k}$  
for each array $\textbf{w}$ of entries $w_{s}^{k}$):
we use
\begin{linenomath*}
\begin{equation}
E(\textbf{q})\equiv f(\textbf{q},\,\textbf{u}_{\mathbf{q}})\equiv \sum_{{\mathcal M}^{(d)}} f_{\overline{s}}^{\overline{k}}\left(\textbf{q},\,\textbf{u}_{\mathbf{q}}\right),
\label{def}
\end{equation}
\end{linenomath*}
to  quantify  
the discrepancy between  data and model. We have set
\begin{linenomath*}
\begin{equation}
\begin{gathered}
f_{\overline{s}}^{\overline{k}}\left(\textbf{q},\,\textbf{u}\right)=
C_{0}^{-2}\left(\left(
B\left(\textbf{q},\textbf{u}\right )\right)_{\bar{s}}^{\bar{k}}-
C_{\bar{s}}^{\bar{k}}\right)^{2},
\end{gathered}
\label{Dis_E}
\end{equation}
\end{linenomath*}
and  $\sum_{{\mathcal M}^{(d)}}$  stands for the complete notation  $\sum_{(\bar{s},\bar{k})\in {\mathcal M}^{(d)}}$. Though $\textbf{B}$, $f$ and  $f_{\overline{s}}^{\overline{k}}$ depend  on  $\theta$ and  
$\textbf{C}$, we do not mention these arguments.

Since  array $\textbf{B}$ is continuously differentiable with respect to $\textbf{q}$  in  $Q_\varepsilon$,  the cost function $E$ has exactly one minimum 
in this closed convex set.  This  minimum characterizes the
parameters  that give  to Eq. (\ref{finalEq}) the best chance of representing mass transport in the experimental conditions where the data stored in $\textbf{C}$ were recorded.
Robust inversion methods  find such a minimum by applying rapidly converging optimization algorithms \cite{Nocedal} which require gradients provided by user.

\section{   Cost function gradient and adjoint state}
\label{AdjStM}

We take advantage of such  algorithms provided we  accurately approximate  the gradient of $E$, which is significantly facilitated
if we use an adjoint state.

\subsection{Adjoint state}
\label{sub:Theoretical-background}

 Indeed, we are searching the minimum of  $f(\textbf{q},\,\textbf{u})$ when $\textbf{u}$ satisfies the linear constraint  (\ref{Dis_sh}).  A standard method of constrained optimization  \cite{Chavent5}
  consists in noticing that for each $\boldsymbol{\psi}$ in $X$ the cost function $E(\textbf{q})$ coincides with  $\mathcal{L}\left(\textbf{q},\,\textbf{u}=\textbf{u}_\textbf{q},\,\boldsymbol{\psi}\right)$ where  
the functional $\mathcal{L}$  defined by
\begin{linenomath*}
\begin{equation}
 \mathcal{L}\left(\textbf{q},\,\textbf{u},\,\boldsymbol{\psi}\right)\equiv  f(\textbf{q},\,\textbf{u})+\langle\left(A\left(\textbf{q},\,\textbf{u}\right)-
  \textbf{r}(\textbf{q})\right)\cdot\boldsymbol{\psi}\rangle_{X}
  \label{Lagr}
\end{equation}
\end{linenomath*}
 depends on the supplementary variable $\boldsymbol{\psi}$.  The latter plays the role of a Lagrange multiplier: far from resulting into a more complex optimization problem, it gives us the opportunity of
 sparing the computation of the $2n+4$ derivatives $\frac{\partial\textbf{u}_{\textbf{q}}}{\partial q_{h}}$ in
\begin{linenomath*}
\begin{equation}
\frac{\partial E}{\partial q_{h}}(\textbf{q})=\frac{\partial\mathcal{L}}{\partial q_{h}}\left(\textbf{q},\,\textbf{u}=\textbf{u}_{\textbf{q}},\,
\boldsymbol{\psi}\right)+  \frac{\partial\mathcal{L}}{\partial\textbf{u}}
\left(\textbf{q},\,\textbf{u}=\textbf{u}_{\textbf{q}},\,\boldsymbol{\psi}\right)
\left( \frac{\partial\textbf{u}_{\textbf{q}}}{\partial q_{h}} \right).
\label{grLagr}
\end{equation}
\end{linenomath*}
Indeed, a clever choice of $\boldsymbol{\psi}$   equates to zero the linear
form $\frac{\partial\mathcal{L}}{\partial\textbf{u}}
\left(\textbf{q},\,\textbf{u},\,\boldsymbol{\psi}\right)$ of $X$. This is easy to see upon re-writing    $\frac{\partial f}{\partial\textbf{u}}\!\left(\!\textbf{q},\textbf{u}\right) (\textbf{w})$ 
as a scalar product $\left \langle\!\frac {\partial f}{\partial\textbf{u}}\!
\left(\!\textbf{q},\textbf{u})\!\cdot\!
  \textbf{w}\!\right)\right\rangle _{\!X} $, so that
\begin{linenomath*}
\begin{equation}\frac{\partial\mathcal{L}}{\partial\textbf{u}}\!
\left(\!\textbf{q},\textbf{u},\boldsymbol{\psi}\right)
  (\textbf{w}) =\left \langle\!\frac {\partial f}{\partial\textbf{u}}\!
\left(\!\textbf{q},\textbf{u})\!\cdot\!
  \textbf{w}\!\right)\right\rangle _{\!X}+ \left \langle A\left(\textbf{q},\, \textbf{w} \right)\cdot\boldsymbol{\psi} \right\rangle_{X}.
\label{droite}
\end{equation}
\end{linenomath*}
The right hand-side of (\ref{droite}) in turn is viewed as a scalar product of the form $\left \langle \textbf{U}\cdot\textbf{w} \right\rangle_{X}$
with the help of operator  $A^{*}$ adjoint to $A$, i.e. satisfying
$\left\langle A(\mathbf{w})\cdot\mathbf{w'}\right\rangle _{X}=\left\langle \mathbf{w}\cdot A^{*}\left(\mathbf{w'}\right)\right\rangle _{X}$
for all $(\mathbf{w},\mathbf{w'}) $ in ${\mathcal D}(A)\times{\mathcal D}(A^{*})$. We
    specify  $A^{*}$ and  ${\mathcal D}(A^{*})$ in Appendix \ref{adjgr} which also shows that
for any  $\textbf{q}$ of  $Q_\varepsilon$ there exists one $\boldsymbol{\psi}_{\textbf{q}}$ in ${\mathcal D}(A^{*})$  solving
\begin{linenomath*}
\begin{equation}
  A^{*}(\textbf{q},{\boldsymbol{\psi}}_{\textbf{q}})=-\frac{\partial f}{\partial\textbf{u}}(\textbf{q},\textbf{u}=\textbf{u}_\textbf{q}),
  \label{Adjequ}
\end{equation}
\end{linenomath*}
  called  adjoint problem of  (\ref{Dis_sh}).
We easily
 deduce  from (\ref{Dis_E}) the right hand-side of (\ref{Adjequ}),
and  solving this problem for $\boldsymbol{\psi}_{\textbf{q}}$ (the adjoint state) is no more difficult that solving the direct equation (\ref{Dis_sh}). Then, simple algebra 
using two technical points detailed in  Appendix  \ref{derss}   gives us 
$\frac{\partial f}{\partial q_{h}}\left(\textbf{q},\,\textbf{u}=\textbf{u}_\textbf{q}\right) $,
 $\frac{\partial A}{\partial q_{h}}(\textbf{q},\textbf{u}=\textbf{u}_\textbf{q})$
and $\frac{\partial \textbf{r}}{\partial q_h}(\textbf{q})$. Thus computing 
the $2n+4$ components of the gradient
\begin{linenomath*}
\begin{equation}\frac{\partial E}{\partial q_{h}}({\textbf{q}})=
\frac{\partial f}{\partial q_{h}}\left(\textbf{q},\,\textbf{u}=\textbf{u}_\textbf{q}\right) 
+\left\langle \left(\frac{\partial A}{\partial q_{h}}(\textbf{q},\textbf{u}=\textbf{u}_\textbf{q})-\frac{\partial \textbf{r}}{\partial q_h}(\textbf{q})\right)\cdot\boldsymbol{\psi}_{\textbf{q}}\right\rangle.
 \label{LE}
\end{equation}
\end{linenomath*}
 spares computational time.
\subsection{Speeding up the optimization loop of Fig.\ref{flowchart}}
\label{action}
Indeed, the optimization process represented on Fig.\ref{flowchart} finds the minimum of $E$ at the end of a sequence of iterations. Each of these needs  updating $E$ and its  gradient  by completing 
   tasks ``Compute  $E(\textbf{q})$'' and  ``Compute  $\nabla E(\textbf{q})$''. The first task 
  solves  the direct  problem (\ref{Dis_sh}) associated with  $\textbf{q}_i$. The second updates
 the gradient of $E$   in three stages : 
\emph{(i)} the right hand-side of the adjoint problem (Eq. (\ref{Adjequ})) is deduced from $\textbf{q}$ and $\textbf{u}_{\mathbf{q}}$- \emph{(ii)}   the adjoint problem is solved for the
 adjoint state $\boldsymbol{\psi}_{\textbf{q}}$-
\emph{(iii)} inserting $\boldsymbol{\psi}_{\textbf{q}}$ and  $\textbf{u}_{\mathbf{q}}$ into   Eq. (\ref{LE}) determines the desired gradient. 
Since stages \emph{(i)} and \emph{(iii)} are of negligible computational cost, we  complete task ``Compute  $\nabla E(\textbf{q})$'' by 
  solving one algebraic system different from (\ref{Dis_sh}) but no more  complex.
   Instead determining the $2n+4$ derivatives
$\frac{\partial\textbf{u}_{\textbf{q}}}{\partial q_{h}}$ by finite differences 
 would require solving this system $2n+4$ times.
Therefore, using $\boldsymbol{\psi}_{\textbf{q}}$  and (\ref{LE}) a priori divides the computing time by at least $n+2$. In fact, we will see in Section \ref{sub:Validation-heterogeneous} 
 that finite differences also waste accuracy, not only time.

\section{Inverse method}
\label{met}
The  loop of Fig.\ref{flowchart} gives us a parameter identification tool for Eq.(\ref{finalEq}): updating the gradient
 of $E$  by adjoint state method   rapidly gives us  the accurate information needed by efficient optimization algorithms satisfying robustness principles described below.
We validate the method in a numerical experiment which also discusses  practical details as tolerance values and interpolation nodes.
\subsection{An algorithm that steps the  $\textbf{q}_i$ sequence}
\label{almet}
 
Several strategies build  $\textbf{q}_i$ sequences that decrease $E(\mathbf{q}_{i})$  to the minimum of the smooth function    $E(\textbf{q})$ \cite{Nocedal}
and avoid trapping in locally flat regions. Among them, efficient quasi-Newton methods determine for $\textbf{q}_{i+1}- \mathbf{q}_{i}$ 
 a direction pointing to the minimum of a convex quadratic local approximation ${\mathcal E}_{i+1}$ 
to $E$, updated at each step to account for the curvature of $E$ observed at most recent step. The modulus of  $\textbf{q}_{i+1}- \mathbf{q}_{i}$ moreover must decrease $E$ without being too short.
The BFGS (Broyden-Fletcher-Goldfarb-Shanno) formula \cite{Nocedal} determines $\textbf{q}_{i+1}- \mathbf{q}_{i}$ according to these principles. It requires the gradient of $E$ at current and previous steps,
but    converges super-linearly to the minimum of the smooth  function $E$. The 
 L-BFGS-B  free software \cite{Byrd} satisfies these requirements, accounts for inequality constraints (as  the definition
 of $Q_\varepsilon$), and has a limited memory version very useful in problems with many degrees of freedom as here.

  Iteratively completing the three  tasks represented  on Fig.\ref{flowchart}   retrieves independent of $x$ parameters   arbitrarily imposed   to numerical  solutions of (\ref{Dis_sh}), with relative error 
smaller than $0.3\%$ \cite{Maryshev13}.
Because using adjoint state in action ``Compute $\nabla E(\mathbf{q})$'' generates economies proportional to the number of freedom degrees, this approach
 is expected more useful when  some  
parameters  depend on space.
We specify the stopping criterion of Fig.\ref{flowchart}  and  check the method efficiency by applying it
on artificial data solving problem (\ref{Dis_sh}), and for which we know the actual value of parameter vector $\mathbf{q}$.

\subsection{Validation, tolerance  and interpolation nodes}  
\label{sub:Validation-heterogeneous} 
\begin{figure}
\begin{center}
\includegraphics[width=0.9\linewidth]{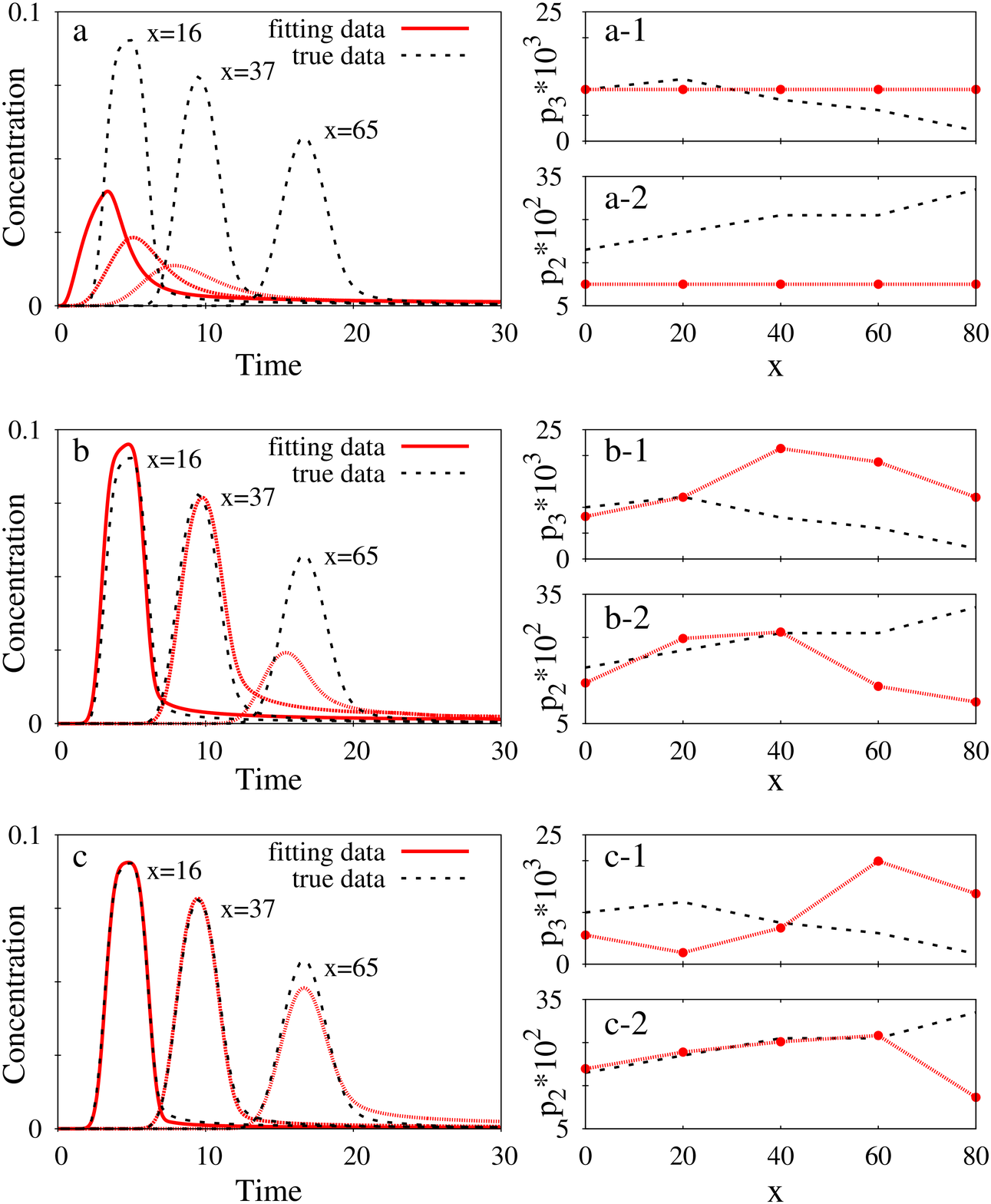}
\caption{ \label{test1}
Profiles  and estimated  space dependent coefficients of Eq.(\ref{finalEq}) at some steps $i$ of the  optimization applied to synthetic data. 
Left: total concentrations. Right: functions $p_2$ and $p_3$. Functions $p_{2,true}$ and $p_{3,true}$ and synthetic data
are in black dashed lines.  Estimated  total concentration   profiles extracted from
$B({\mathbf{q}}_i,\textbf{u}_{{\mathbf{q}}_i})$  and tentative estimates of $p_2(x)$ and $p_3(x)$ deduced from ${\mathbf{q}}_i$ are in red full lines.
Step number $i$  is documented in Table 1.}
\end{center}
\end{figure}

\begin{figure}
\begin{center}
\includegraphics[width=0.9\linewidth]{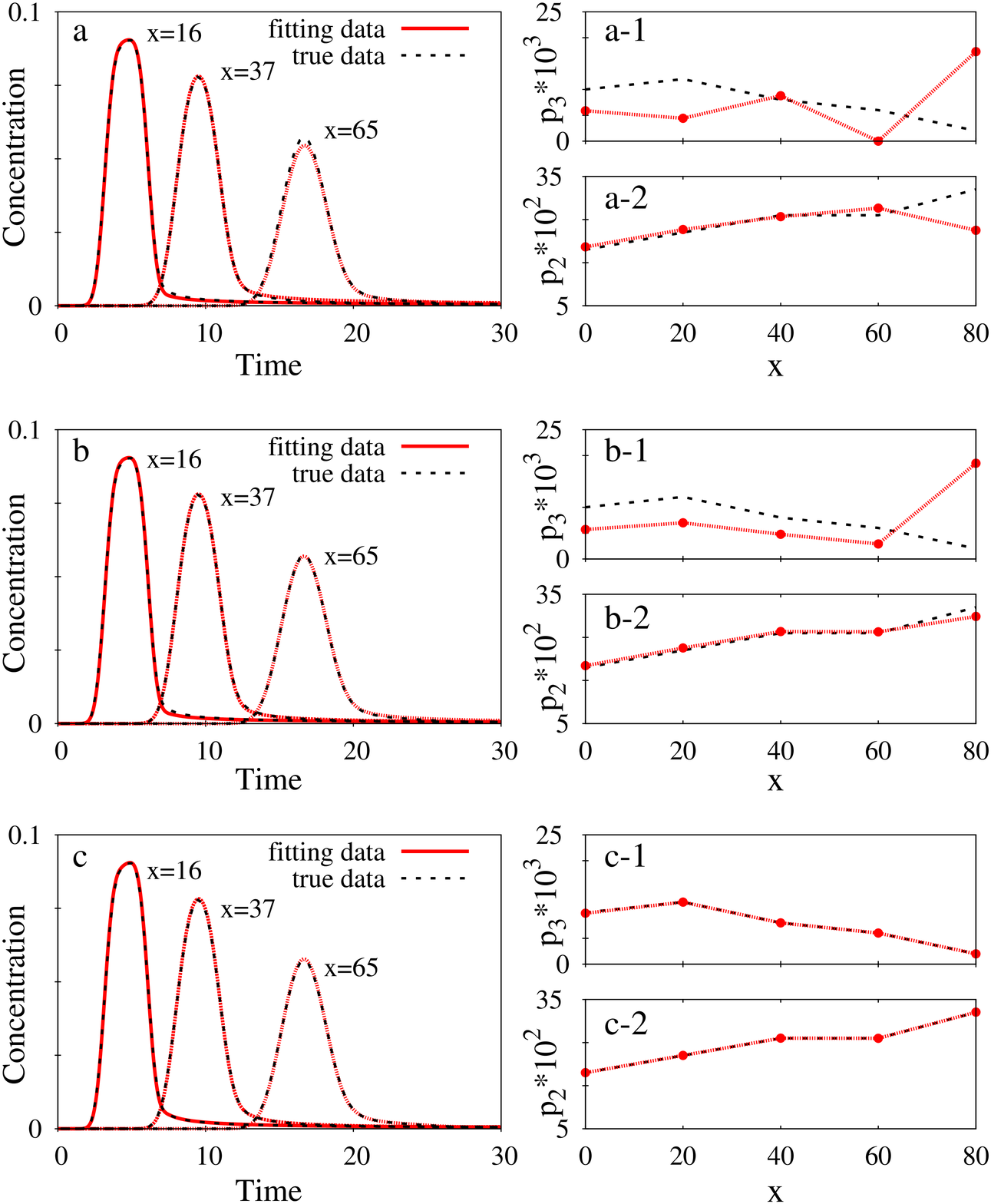}
\caption{\label{test2} 
Profiles  and estimated  coefficients of (\ref{finalEq}) at several steps of the  optimization applied to synthetic data, continued.
 Optimization is stopped at step ($i_f=1100$) represented on c.
}
\end{center}
\end{figure}

We  construct  such data  by solving the
 discrete direct problem (\ref{Dis_sh})  associated to arbitrary  functions $p_{j,true}(x)$ for $j=2,3$ and  numbers  $p_{1,true}$  and $\alpha_{true}$. 
Function $\theta(x)$ is also
 arbitrarily  chosen. In the example discussed immediately below,  $\theta(x)=0.2\times(1+x/L)$,  $\alpha_{true}=0.7$ and $p_{1,true}=10^{-2}$ cm$^{2}$/h. The piecewise linear functions $p_{2,true}$ and  
$p_{3,true}$ are  represented by black dashed lines  at the right of Figures \ref{test1} and \ref{test2}.
We store $p_{1,true}$, the interpolation values of  $p_{2,true}$ and  $p_{3,true}$ and $\alpha_{true}$ in   parameter vector $\mathbf{q}_{true}$. 
 Then, inserting in Eq.(\ref{totCdis})  the solution of Eq.(\ref{Dis_sh}) gives us time profiles of  $\textbf{B}(\mathbf{q}_{true},\mathbf{u}_{\mathbf{q}_{true}})$.  
 For three values of $x$, this quantity is represented by three black dashed lines on all graphs at the 
left of Figures \ref{test1} and \ref{test2}. These profiles play the role of the data $\mathbf{C}$ that define the objective
 function $E$ in Eqs.(\ref{def}-\ref{Dis_E}). We imagine that we do not know the true parameters and  estimate them by applying to  $\mathbf{C}$
 the optimization process described 
 in Section \ref{almet}. But before, we must choose the $n+1$ interpolation nodes. This determines the dimension of $Q_\epsilon$. We fix the nodes
 by trial and error, beginning with small  $n$.  
With the very simple  functions  $p_{2,true}$ and  $p_{3,true}$ of Figs. \ref{test1}-\ref{test2}, $n=4$ is sufficient. However, more 
complex coefficients  need
larger values.

\begin{figure}
\begin{center}\includegraphics[width=0.9\linewidth]{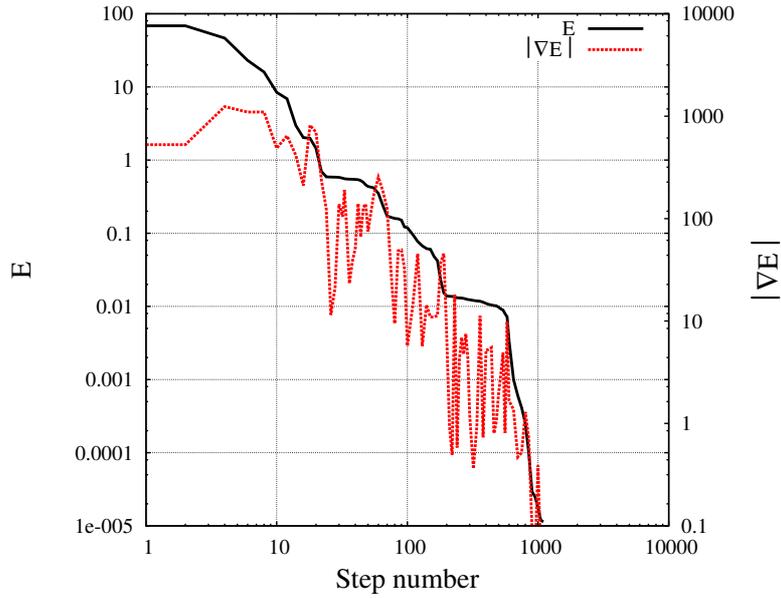}
\caption{\label{FCPE_grad}  Cost
function  $E({\mathbf{q}}_{i})$ and the modulus of its gradient along the optimization procedure applied to the synthetic data represented on Figs \ref{test1}-\ref{test2}. 
  The  $E({\mathbf{q}}_{i})$  sequence exhibits subsets where it varies slowly, but this is because the $\mathbf{q}_{i}$ sequence itself steps slowly: the  too large gradient
 reveals that $E({\mathbf{q}})$ is far from stationary there.
Continuing the process in such case is justified a posteriori by Table \ref{tab1} and   Fig.\ref{test2}.
}
\end{center}
\end{figure}

For each of these choices  we fix  $\mathbf{q}_0$ at random, and  stop the sequence when  $ E({\mathbf{q}_i})$ becomes stationary provided 
 $|\nabla E({\mathbf{q}})|$ also stabilizes without being too large.
  Since we know the true parameters,  the  numerical experiment gives 
 us the opportunity of discussing a limit tolerance value for $|\nabla E|$ at final estimate $\mathbf{q}_{i_f}$ on the basis of  Fig.\ref{FCPE_grad}.  We observe the decrease 
 of $E(\mathbf{q}_i)$ which takes almost
  stationary  values  
for $i$ in $\left\{200,...,800\right\}$. Yet, they correspond to non-small values of $|\nabla E({\mathbf{q}_{i}})|$ which suggest that the true minimum of
 $E$ ($0$ in this case) is not observed yet. 
  Table 1 and Figs \ref{test1}-\ref{test2} 
confirm that the corresponding  $\mathbf{q}_i$  are poor estimates of $\mathbf{q}_{true}$:
tolerance values of $|\nabla E({\mathbf{q}})|$  above $10^{-1}$ are too large for the problem at hand. Yet there is no general rule, 
and we are ready to try tolerance values as small as  possible. In fact,
  continuing the numerical experiment
 further does not improve the estimate. 
\begin{table}
\begin{center}
\begin{tabular}{|c|c|c|c|c|c|c|c|}
\hline 
 {plot} &  {Fig. \ref{test1}a}  & {Fig. \ref{test1}b}
 & {Fig. \ref{test1}c}& {Fig. \ref{test2}a} &{Fig. \ref{test2}b} &{Fig. \ref{test2}c} \tabularnewline
\hline 
 {  $p_1, cm^2/h$ } & {$3$}  & {$2\times 10^{-7}$} &  { $0.0206$}& {$0.0223$}& {$0.0228$} & {$0.0102$}\tabularnewline
\hline 
{ $\alpha$} & { $0.5$} & {$0.529 $ }  & { $ 0.532$}& {$0.53$}& {$0.547$}& {$0.697$}\tabularnewline
\hline 
 { $E({\mathbf{q}}_i)$ } &  {$70$}  & {$8.4$} &  { $0.54$} & {$6\times 10^{-2}$}& {$10^{-2}$}& {$ 10^{-5}$}\tabularnewline
\hline
 { Step No $i$ } &  {$0$}  & {$10$} &  { $40$} & {$140$}& {$400$}& {$ 1100$}\tabularnewline
\hline 
 
\end{tabular}
\caption{ \label{tab1} Estimated space independent parameters and cost function at some steps of   optimization process applied to artificial data represented on  Figures \ref{test1} and \ref{test2}. 
Estimates documented in 
the table correspond to optimization steps   represented   on the figures.
  }
\end{center} 
\end{table}

We also tried to conduct the  numerical experiment with  $\nabla E(\mathbf{q})$ computed by finite differences instead of adjoint state.  
  Approximate gradients of  $E$ deduced from (\ref{LE})  coincide with  
  finite difference $\frac{ E(\mathbf{q}+\delta \mathbf{q})-E(\mathbf{q})}{|\delta \mathbf{q}|}$ of small enough 
increments $\delta \mathbf{q}$ \cite{Chavent5}. Numerical comparisons \cite{Maryshev13} confirm that for each $\eta>0$ and each $h=1,...,2n+4$ there exists $\eta'(\eta,\mathbf{q},h)$
 such that  $|\delta \mathbf{q}|<\eta'$ implies
$|\frac{\partial E(\mathbf{q})}{\partial q_h}-\frac{ E(\mathbf{q}+\delta \mathbf{q})-E(\mathbf{q})}{|\delta \mathbf{q}|}|<\eta$ when  vector $\delta \mathbf{q}$  has all
 its entries
 equal to zero except that of rank $h$. Nevertheless, $\eta'$ depends on $\mathbf{q}$: we cannot guarantee any general $\eta'$ valid during the entire
optimization process. Therefore, accurately computing gradients with finite differences needs too many checks to confirm accuracy. 
Implementing these  checks  in an automatic process is too heavy, and forgetting them returns too poor accuracy.
   We experienced this  by running the  numerical experiment with gradient  deduced  from finite differences of very small but fixed  value. For some $\mathbf{q}$ this $|\delta \mathbf{q}|$
 was not small enough and the gradient was not accurate. This resulted into extremely poor final value of $E$ (of $0.1$ compared with the  better result $10^{-5}$ of Table 1) after sixty hours 
(four with adjoint state).  

 These arguments predict  that adjoint state method   will be even more useful with  actual experimental data associated to parameters strongly suspected to vary in space.

\section{Inverting  actual experimental data}
\label{App}
  Actual experimental data  require preliminary  processing  and technical choices.

\subsection{Technical preliminaries}
\label{choix}

 The left panel of Fig.\ref{WC_exp}  displays some of the  $15$  total concentration profiles  recorded by DXS in the device of Fig.\ref{exp}. The $15$ profiles  collect an amount of $3218$ triples $(x,t,C(x,t))$ among which ${\mathcal N}^{(n)}=357$ 
 exhibit  negative  $C(x,t)$ records  revealing measurement errors. We exclude  from set ${\mathcal M}^{(d)}$ the indexes of all items that are negative or observed after negative records in 
  descending slope (or before negative records in ascending slope). 
  It then remains ${\mathcal N}^{(d)}=2861$ non-zero elements in 
the array $\textbf{C}$ that defines
$E$ in (\ref{def}-\ref{Dis_E}).

\begin{figure}
\begin{center}\includegraphics[width=0.8\linewidth]{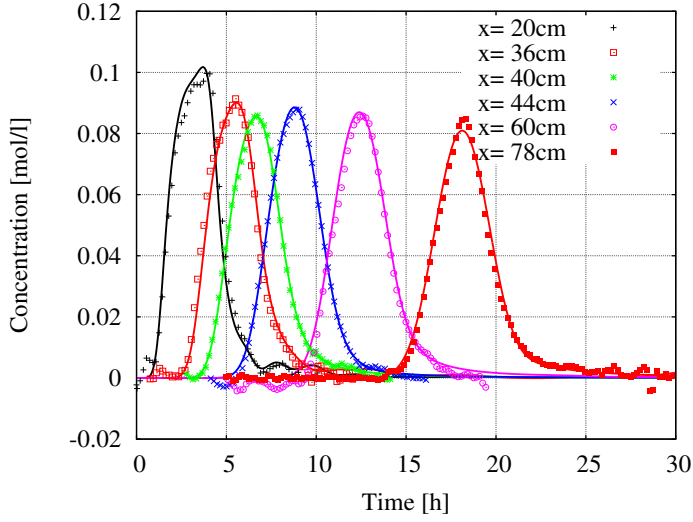}
\caption{\label{Cprof} Total concentration, measured or deduced from $B({\mathbf{q}}_{i_f},\textbf{u}_{{\mathbf{q}}_{i_f}})$.
 Dots and lines 
represent experimental data
 and $B({\mathbf{q}}_{i_f},\textbf{u}_{{\mathbf{q}}_{i_f}})$. 
Fitted values of parameters assumed independent of $x$ are $p_1=3.5\times10^{-3}$ cm$^{2}$/h, $\alpha=0.89$. The estimated mobile water
content profile $p_2\left(x\right)$ is represented on Fig. \ref{parProf}.}
\end{center}
\end{figure}

Before processing these data, we  fix  the numerical mesh and the  details of the  interpolation of $\theta$, $p_2$ and $p_3$.
Space-time step lengths $\Delta x$ and $\Delta t$ common to discrete   problems (\ref{Dis_sh}) and (\ref{Adjequ}) are fixed according  to the order of magnitude  suggested by Appendix \ref{Valid_fMIM} for $N_{Sp}$, 
taking care 
that the lower limit   $V\frac{\Delta t}{\Delta  x}$ of $p_2$  in the definition of $Q_\varepsilon$ does not exclude physically relevant tentative values: with $V=1.05$ cm/h,
 $\Delta x= 0.25cm$ and  $\Delta t= 6\times 10^{-3}h$  exclude  $p_2$ values
smaller than 
 $0.05$, i.e   smaller than   useful values. A posteriori comparisons to histograms of random walks  \cite{Schumer,Benson,NeelZ} approaching the solutions of (\ref{finalEq})
 as in \cite{Ouloin} confirm that these  step
 lengths are small enough.
In addition to the choice of the $n$ base points for the interpolation of $p_2$ and $p_3$  discussed in Section \ref{sub:Validation-heterogeneous}, we also interpolate the measured water content $\theta$ 
because of the high dispersion observed on this quantity. We  use cubic splines with  $n_\theta$  base points, taking care that interpolation coincides with local averages. Proceeding by trial and error, we 
 progressively increase 
 $n_\theta$ and $n$. The 
agreement between $B({\mathbf{q}},\textbf{u}_{{\mathbf{q}}})$ and the data represented  on Fig.\ref{Cprof} is achieved with $n_\theta=18$ and $n=31$ nodes, less spaced in the first half of the column
 (where they are $21$) than for $x$ between $40cm$ and $80cm$.

\subsection{Calibration of experimental data}
\label{sub:Calibration-of-experimental}
Each of these choices determines one minimizing sequence which follows its course automatically according to Fig.\ref{flowchart}, and  we stop it  at step $i_f$ when the 
$E({\mathbf{q}}_{i})$ sequence
ceases moving  provided the
  gradient of $E$ satisfies
$|\nabla E({\mathbf{q}}_{i})| < 10^{-1} $ as suggested in Section \ref{sub:Validation-heterogeneous}. Estimates of  $\alpha$ and  $p_1$ then are $0.90$ and  $3.5\times 10^{-3}cm^2/h$, and   normalized relative error 
\begin{linenomath*}$$e_{\!R}\!=\!\frac{C_0\sqrt{ E({\mathbf{q}}_{i_f})}}{\sum_{\bar{s},\bar{k}}C(\bar{s}\Delta x,\bar{k}\Delta t)} \!\approx\! 2\times 10^{-3} $$
\end{linenomath*}
is about one order of magnitude larger than with artificial data (in Section \ref{sub:Validation-heterogeneous}). The averaged  absolute deviation from the data is
\begin{linenomath*}$$ e_A=C_0\sqrt{\frac{ E({\mathbf{q}}_{i_f})}{{\mathcal N}^{(d)}}}=3.4\times 10^{-3}mol/l.$$
\end{linenomath*}

It is about $3$ times the   measurement error lower bound $\sqrt{\frac{\sum_{{\mathcal M}^{(n)}}C(x,t)^2}{{\mathcal N}^{(n)}}}=1.4\times 10^{-3}mol/l$ suggested by the negative 
 concentration records  excluded from array $\mathbf{C}$
in section \ref{choix}, set ${\mathcal M}^{(n)}$ representing the corresponding indexes. Observing the same magnitude order  for  $e_A$ and  measurement error lower bound suggests that 
the here considered data  are not  badly represented by Eq. (\ref{finalEq}), here associated with small but non-negligible trapping time heterogeneity manifested by 
the estimate of $\alpha$. 

The  estimated value of $p_1$ suggests that we use a finite difference scheme  flawed by numerical diffusion. Hence, we compared $B({\mathbf{q}}_{i_f},\textbf{u}_{{\mathbf{q}}_{i_f}})$
with the issue of the smart version of the discrete direct problem briefly described at the end of Appendix \ref{Valid_fMIM}. Numerical dispersion is observed. Nevertheless, the discrepancy between the two schemes
is negligible in comparison with $e_{\!R}$. Also remind that we assumed parameter $p_1$ independent of $x$. In fact, relaxing this assumption returns the same estimates because $p_1$ is small: 
we are in a regime where
the solutions of (\ref{finalEq}) are sensitive to the order of magnitude of this parameter, not to its local variations.
 
\begin{figure}
\begin{center}\includegraphics[width=0.9\linewidth]{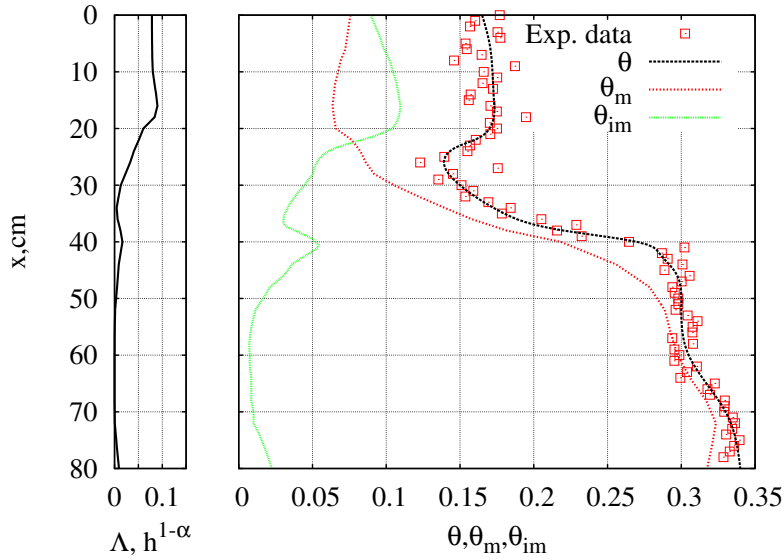}
\caption{\label{parProf} Profiles of space-dependent parameters $p_2=\theta_m$, $\Lambda=p_3/p_2$ and $\theta_{im}=\theta-p_2$. Parameters  $\theta_{m}\left(x\right)$, $\theta_{im}\left(x\right)$ and
$\Lambda\left(x\right)$ are estimation results,   $\theta\left(x\right)$ is measured (dots) and  interpolated (black line).}
\end{center}
\end{figure}

 Estimated profiles 
of $\theta_m=p_2$  and $\Lambda=p_3/p_2$ deduced from final ${\mathbf{q}}_{i}$ are represented on Fig.\ref{parProf}, along with 
 $\theta_{im}=\theta-p_2$. The latter and $\Lambda$ show similar slopes in  agreement with the assumption that $\Lambda$  is nearly proportional to
 the immobile fluid mass as the mass-exchange coefficient of the standard MIM \cite{Genuch}. 

\section{Conclusion}

Adjoint  states designate solutions of very diverse  equations
 \cite{Sun,Neupauer99,Maryshev13,Sykes,SunYeh90,Neupauer04}
 including  linear operators adjoint to the left hand-side of a p.d.e. as  (\ref{finalEq}), or to a discrete formulation .
 The second  possibility is the most efficient for parameter identification \cite{SanduG,WilsonT}  minimizing the distance $E$  between data and  numerical
solutions to  any p.d.e.

 Such a minimum gives us the coefficients of
Eq. (\ref{finalEq}) and the  order $\alpha$ of the fractional derivative  the best adapted to   dispersion data composed of  a series of  solute concentration  time profiles
recorded at several locations of a medium.   Heat transfer data were previously processed with a time fractional diffusion  p.d.e, 
yet with much less degrees of freedom \cite{Ghazizadeh} than here because we account for space dependent coefficients  (necessitating 
sixty degrees of freedom).

 The task is feasible though the  many degrees of freedom because we do not compute  the  derivatives of $\textbf{u}_\textbf{q}$ w.r.t. the components  $q_h$ of $\textbf{q}$, and
 re-formulate
 the cost 
function $E$ by introducing  one adjoint state
 that cancels the influence  of these  derivatives. Instead of solving as many  discrete copies of Eq.(\ref{finalEq})  as there are degrees of freedom,
   we  determine \cite{Chavent5}
this adjoint state that solves one adjoint problem of the same complexity as
 the direct problem that gives us  $\textbf{u}_\textbf{q}$. This allows us varying necessary arbitrary choices which are known to strongly influence the optimization issue.

 \section*{Acknowledgements}
The authors have been supported  by Agence Nationale de la Recherche (ANR project ANR-09-SYSC-015), B. Maryshev being postdoctoral fellow at 
DEN-DANS-DM2S-STMF-LATF CEA/Saclay in 2011-2013.

\appendix

\section{A discrete version of the fractional MIM}
\label{Discr} 
  Standard  approximations to
 fractional integrals and  derivatives  define the discrete  problem (\ref{Dis_sh}) whose solutions  approximate those  of (\ref{finalEq}).

\subsection{Approximating temporal  integrals and derivatives} 
\label{Discrint} 
In  $p_2\partial_t+p_3\partial_tI_{0,+}^{1-\alpha}$ and in (\ref{totC})  the Riemann-Liouville integral  $I_{0,+}^{1-\alpha}$ accounts for past history since  time $t=0$.
An approximation of  order $O\left(\Delta t^{2}\right)$ is \cite{Diethelm}
\begin{linenomath*}
\begin{equation}
\begin{gathered}
I_{0,+}^{1-\alpha}y(k\Delta t)\approx \sum_{j=0}^{k}{\mathcal I}^{j, k}y^{k-j}\quad  \textrm{with} \quad{\mathcal I}^{0, k}= {\mathcal I}^0\equiv\frac{\Delta t^{1-\alpha}}{\Gamma\left(3-\alpha\right)},\\
  {\mathcal I}^{j, k}={\mathcal I}^{0}[(j+1)^{2-\alpha}-2j^{2-\alpha}+(j-1)^{2-\alpha}]\: \textmd{for}\: 0<j<k, \\
{\mathcal I}^{0,0}=0 \:\textrm{and }\:{\mathcal I}^{k, k}={\mathcal I}^{0}[(2-\alpha)k^{1-\alpha}-k^{2-\alpha}+(k-1)^{2-\alpha}] \:\textmd{ for}\: 0<k.
\end{gathered}
\label{iq}
\end{equation}
\end{linenomath*}
 Combining (\ref{iq}) with the  standard backward
finite difference approximation  $\frac{y^{k}-y^{k-1}}{\Delta t}$ to the first order derivative yields 
\begin{linenomath*}
\begin{equation}(p_2\partial_t+p_3\partial_tI_{0,+}^{1-\alpha})y(k\Delta t)= \frac{1}{\Delta t}\sum_{j=0}^kW_s\left(j,\, k\right)y((k-j)\Delta t)+O(\Delta t),
\label{A4}
\end{equation}
\end{linenomath*} with
\begin{linenomath*}
\begin{equation}
\begin{gathered}
W_s\left(0,\, k\right)=
    p_{2}+p_{3}{\mathcal I}^0 \quad\textrm{and} \\
W_s\left(j,\, k\right)= p_{3} \left[{\mathcal I}^{j, k}-{\mathcal I}^{j-1, k-1}\right]+p_{2}\delta_{j,1}\:\textmd{\small for}\: 1\leq j\leq k.
\end{gathered}
\label{Weq}
\end{equation}
\end{linenomath*}

\subsection{Approximating spatial derivatives and boundary conditions}\label{spe} 

 At each time step $k>0$ and for each $s\in\left\{1,...,N_{Sp}\right\}$,  standard 
central finite differences 
\begin{linenomath*}
\begin{equation}
\frac{p_1}{\Delta x^{2}}\left[ -\!u_{s+\!1}^{k}\!+\!\!2u_{s}^{k}-u_{s\!-\!1}^{k\!}\right] +\frac{V}{2\Delta x}\left[u_{s+\!1}^{k}\!-\!u_{s\!-\!1}^{k\!}\right],
  \label{sp}
\end{equation}
\end{linenomath*}
  approximate $-\partial_{x}\left(\partial_{x}p_1u-Vu\right)|_{s\Delta x}^{k\Delta t}$  at order $O(\Delta x^2)$. Applying non-centered finite differences to the  boundary conditions  in (\ref{contibc})  links
 $u_{0}^{k}$ 
and $u_{N_{Sp}+1}^{k}$  to immediately neighboring $u_{s}^{k}$, but is accurate at first order only: 
\begin{linenomath*}
\begin{equation}\label{bou}
{\Delta xV+p_1}u_{0}^{k}= {p_1u_{1}^{k}}+{C_0V}{\Delta xV+p_1}H(t_f-k\Delta t)+O(\Delta x)\: , u_{N_{Sp}+1}^{k}=u_{N_{Sp}}^{k}+O(\Delta x).
\end{equation}
\end{linenomath*}
Hence  equations (\ref{A4}) and (\ref{sp})  yield  approximations to 
$$-\partial_{x}\left(\partial_{x}p_1u-Vu\right)|_{s\Delta x}^{k\Delta t}+[(p_2\partial_t+p_3\partial_tI_{0,+}^{1-\alpha})u]|_{s\Delta x}^{k\Delta t}$$ 
at order $O(\Delta x)+O(\Delta t)$
for each  $s\in\left\{1,...,N_{Sp}\right\}$ representing an interior point of $[0,\ell]$ and each index $k>0$.   Equating them to zero and eliminating $u_{0}^{k}$ and
 $u_{N_{Sp}+1}^{k}$
with the help  of  (\ref{bou}) yields a system of equations  determining the   $u_s^k$ that approximate the $u(s\Delta x, k\Delta t)$ at interior points of $[0,\ell]$.
It accounts for solute injection at column inlet. Remembering the initial condition, we set these equations  in the compact form (\ref{Dis_sh}) reproduced below
\begin{linenomath*}
\begin{equation}
  A(\textbf{q},\textbf{u})=\textbf{r}(\textbf{q}), \quad\textbf{u}\in {\mathcal D}( A)\nonumber
\end{equation}
\end{linenomath*}
by defining  linear mapping $A$  and  array $\textbf{r}$.  Each array $\textbf{u}$ of
 ${\mathcal D}( A)\equiv\left\{\mathbf{u}\in X/\mathbf{u}^{0}=0\right\}$ satisfies  the initial condition included in (\ref{contibc}) and $A$ maps it  onto array $A(\mathbf{q},
{\textbf u})$ whose each rank $k$ column $(A(\mathbf{q},
{\textbf u}))^{k}$ is 
\begin{linenomath*}
\begin{equation}
(A(\mathbf{q},
{\textbf u}))^{k}\equiv  {\textbf G}\textbf{u}^{k}+\sum_{j=1}^{k} \textbf{W}\left(j,\,k\right)
\textbf{u}^{k-j}, 
  \label{Dis_A}
\end{equation}
\end{linenomath*}
where  $\textbf{u}^k=(u_{1}^{k},...,u_{N_{Sp}}^{k})^\dag$ represents the rank $k$ column of $\textbf{u}$. Matrices ${\textbf G}$ and $ \textbf{W}\left(j,\,k\right)$ are defined at the end of the section, and
array $\textbf{r}$ recollects the $r_s^k$ defined by
\begin{linenomath*}
\begin{equation}
   r_{1}^{k}=\Delta t\frac{C_{0}V}{2\Delta x}\frac{2p_1+V\Delta x}{V\Delta x+p_1}H(k\Delta t)H(t_{f}-k\Delta t), \:\: r_{s}^{k}=0\:\: \textrm{for}\:\: s>1 \:\:\textrm{or}\:\: k=0.
   \label{Source}
\end{equation}
\end{linenomath*}
 Each rank $k$  column $(r_{1}^{k},...,r_{N_{Sp}}^{k})^\dag$ of $\textbf{r}$ being noted $\textbf{r}^k$, (\ref{Dis_sh}) is equivalent to the system of equations
\begin{linenomath*}
\begin{equation}
(A(\mathbf{q},
{\textbf u}))^{k}=\textbf{r}^{k}\quad \textrm{\small for}\quad 0\leq k\leq N_T.
  \label{Dis_eq}
\end{equation}
\end{linenomath*}
With $\mu=\frac{p_1\Delta t}{\Delta x^{2}}$ and
$\nu=\frac{V\Delta t}{2\Delta x}$,  the entries $g_{s,s'}$ of $\textbf{G}$ are 
\begin{linenomath*}
\begin{equation}
\begin{gathered}
g_{s,s'}=0\quad \textmd{\small for}\quad |s-s'|>1,\quad g_{s,s-1}=-\nu-\mu \quad\textmd{\small for}\quad s>1,\\
   g_{s,s+1}=\nu-\mu \quad \textmd{\small for}\quad s<N_{Sp}, \quad
 g_{s,s}= p_{2}(s\Delta x)+p_{3}(s\Delta x) {\mathcal I}^0+\Omega_s,
\end{gathered}
\label{Adiscr}
\end{equation}
\end{linenomath*}
where all interior diagonal entries exhibit the same $\Omega_s=2\mu $ (for $ s\neq1$ and $s\neq N_{Sp}$), while boundary conditions (\ref{bou}) result into 
 $ \Omega_1=\mu(1+\frac{1}{2+\mu/\nu})$ and
$\Omega_{N_{Sp}}=\mu+\nu$ at both ends.
We see that matrix $\textbf{G}$ depends on $\textbf{q}$, $\Delta x$ and $\Delta t$.
 The $\textbf{W}\left(j,\, k\right)$ are  $N_{Sp}\times N_{Sp}$
 diagonal matrices of entries
$(\textbf{W}\left(j,\, k\right))_{s,s}=W_s\left(j,\, k\right)$. In (\ref{Dis_A}) each $\textbf{W}\left(j,\, k\right)$  operates on columns  of  rank smaller than $k$ stored  in  $\textbf{u}$.

 Using (\ref{Dis_A}) will help us seeing that 
 problem (\ref{Dis_sh}) has exactly one solution in ${\mathcal D}( A)$
provided these parameters make matrix  $\textbf{G}$ invertible.

\subsection{Well-posedness of  (\ref{Dis_sh})}\label{well}%
\label{Diag_dom}
If $\textbf{G}$ is invertible, we  easily determine the single solution $\mathbf{u}_\mathbf{q}$  of  (\ref{Dis_eq}) in  ${\mathcal D}( A)$    by setting  ${\textbf u}^{0}=0$ and  recursively solving  equations  (\ref{Dis_eq}), increasing  $k$ from 
$1$ to $N_T$. 
Gershgorin circle theorem 
\cite{Gersh}  gives us a sufficient condition for   matrix $\textbf{G}$ invertibility in the form of
\begin{linenomath*}
\begin{equation}
  \left|g_{s,s}\right|>\left|g_{s-1,s}\right|(1-\delta_{s,1})+\left|g_{s+1,s}\right|(1-\delta_{s,N_{Sp}}),
  \label{DiagCond}
\end{equation}
\end{linenomath*}
 $\delta_{i,j}$ being Kronecker index.
This condition is 
 satisfied when $ {\textbf q}$ belongs to any  closed convex set  $ {Q}_\varepsilon= {R}_+\times[\frac{V\Delta t}{\Delta x}+\varepsilon,+\infty[^{n+1}\times  {R}_+^{n+1}\times[0,1]$ with $\varepsilon>0$
arbitrarily small.

 By Implicit Function Theorem \cite{Jittorn} the mapping  $\textbf{q}\mapsto\textbf{u}_{\mathbf{q}}$   is
of ${\mathcal C}^\infty$ class in   
$Q_\varepsilon$,   matrices $\textbf{G}$ and  $\textbf{W}\left(j,\, k\right)$ having elements  of class ${\mathcal C}^\infty $ in  
$ {Q}_\varepsilon$ with respect to the components
of  $ {\textbf q}$.

\subsection{Numerical scheme accuracy}

\label{Valid_fMIM}
 We validate the above scheme and fix  $\Delta x$ and $\Delta t$ by considering  a    ${\mathcal R}\neq 0$ so that the continuous problem   (\ref{finalEq}-\ref{contibc}) with $C_0=0$ has an exact solution which we
 compare with the still noted  $u_s^k$  issues of the  above scheme.
We set
\begin{linenomath*}$$\varphi_0(x)=e^{\frac{Vx}{2p_1}}(\sin{\sigma x}+\frac{2p_1\sigma}{V}\cos{\sigma x})$$
\end{linenomath*}
 and take $\sigma$ defined by
\begin{linenomath*}$$\frac{2p_1}{V}\sigma^{2}\sin{(\sigma L)}-2\sigma\cos{(\sigma L)}-\frac{V}{2p_1}\sin{(\sigma L)}=0:$$ 
\end{linenomath*}
  $u(x,t)\equiv t\varphi_0(x)$ solves  (\ref{finalEq},\ref{contibc})
 provided we set
\begin{linenomath*}$${\mathcal R}(x,t)=\varphi_0(x)\left[p_2(x)+p_3(x)\frac{t^{1-\alpha}}
{\Gamma(2-\alpha)}+Mt\right],\quad  M=p_1\sigma^{2}+\frac{V^{2}}{4p_1}.$$ 
\end{linenomath*}
Steps $\Delta x$ and $\Delta t$   must satisfy $p_{2}(s\Delta x)> \frac{V\Delta t}{\Delta x}$ to ensure $\textbf{q}\in Q_\varepsilon$.
 For instance with   $L=1$, $p_1=0.05$, $V=1$, $\alpha=0.8$, $p_2(x)=0.5+0.3\sin{(4\pi x})$ and $p_3(x)=0.5-0.4\sin{(4\pi x)}$, taking $N_{Sp}=360$ and 
$\Delta t=\Delta x/10$ results into relative errors
$\left |\frac{u_s^k-u(s\Delta x,k\Delta t)}{u(s\Delta x,k\Delta t)} \right |$ smaller than $2\times 10^{-4}$, for $t<1$. 

This gives us an idea of which grid  we can use.
A posteriori  checks are applied, especially when parameter identification returns large P\'eclet numbers. In this case  the centered finite difference approximation used in Appendix
\ref{spe} for the advection term $V\partial_xu$ may be flawed by numerical diffusion. Hence, we compare with a modified version of the discrete direct problem
  (\ref{Dis_sh}) in which
$\frac{V}{6\Delta x}(3u_s^n+u_{s-2}^n-6u_{s-1}^n+2u_{s+1}^n)$  approximates  $V\partial_xu$. This is Eq. (4.9) of
 \cite{Torrilhon} which corresponds to a flux limiter given by the first line of (3.8) in this reference, in view of the small $\partial^2_{x^2}u$ which we observe.
 Other a posteriori checks are comparisons with random walks whose probability density function is $p_2u$, as in \cite{Ouloin}.
\section{Discrete adjoint state for discrete direct  problem (\ref{Dis_sh})}
\label{adjgr}

The linear operator $A$ is described  by (\ref{Dis_A}), which implies 
\begin{linenomath*}
\begin{equation}
\left \langle A(\textbf{q},\textbf{u})\!\cdot\!\textbf{w} 
\right \rangle_{\!X}\!=\!\!
  \sum_{k=1}^{N_{T}} \!\left \langle \textbf{G}\textbf{u}^{k}\!\cdot\!
\textbf{w}^{k}\right\rangle_{ {R}^{N_{\!Sp}}}\!\!+\!\!
  \sum_{k=1}^{N_{T}}\sum_{j=1}^{k} \left\langle\textbf{W}(j,\, k)
\textbf{u}^{k-j}\!\cdot\!\textbf{w}^{k}\right\rangle_{ {R}^{N_{Sp}}},
\label{eq:Adjoint_DisMIM}
\end{equation}
\end{linenomath*}
for each  $(\textbf{u},\textbf{w})$ in  ${\mathcal D}(A)\times X$. Matrices  $\textbf{G}$ and  $\textbf{W}(j,\, k)$ are defined in  Appendix \ref{spe}, and each $\left \langle \textbf{G}\textbf{u}^{k}\!\cdot\!
\textbf{w}^{k}\right\rangle_{ {R}^{N_{\!Sp}}}$ is equal to $\langle\textbf{u}^{k}\cdot \textbf{G}^{\dagger} \textbf{w}  ^{k}\rangle_{ {R}^{N_{Sp}}}$,  superscript ${\dagger}$ denoting transpose.
 Re-arranging the double sum   on the right hand-side of (\ref{eq:Adjoint_DisMIM}) 
 proves that the adjoint
 $A^{*}$ of $A$ is the operator that   transforms each array $\textbf{w}$ of $X$  into array $A^{*}(\textbf{q},\textbf{w})$ whose each column of rank $k$ is
\begin{linenomath*}
\begin{equation}
(A^{*}(\textbf{q},\textbf{w}))^k=\textbf{G}^{\dagger}\textbf{w}^{k}+
  \sum_{j=1}^{N_{T}-k}\textbf{W}\left(j,\, k+j\right)
\textbf{w}^{k+j}\quad{\mbox {for}} \quad 0\leq k\leq N_T.
\label{discradj}
\end{equation}
\end{linenomath*}
 This implies that Eq.(\ref{Adjequ}) 
 is equivalent to the set of  all equations 
\begin{linenomath*}
\begin{equation}
 \textbf{G}^{\dagger}\boldsymbol{\psi}^{k}+\sum_{j=1}^{N_{T}-k}\textbf{W}\left(j,\, k+j\right)\boldsymbol{\psi}^{k+j}=-\sum_{{\mathcal M}^{(d)}}\left (\frac{\partial f_{\bar{s}}^{\bar{k}}}{\partial\mathbf{u}} \right )^{k} 
(\textbf{q},\,\textbf{u}_{\mathbf{q}})
  \label{discradjpart}
\end{equation}
\end{linenomath*}
 obtained for $0\leq k\leq N_T$. When $\textbf{G}$ is invertible it is the same for $ \textbf{G}^{\dagger}$, hence the  system  of all these equations has exactly one solution  in 
${\mathcal D}(A^*)=\left\{\textbf{u}\in X \big / \textbf{u}^{N_T}=0\right\}$ because 
 the final simulation time $T$ is such that 
$\max\left\{\bar{k} \big / (\bar{k},\bar{s})\in
{\mathcal M}^{(d)}\right\}<N_{T}$ which implies
 $\left(\frac{\partial f_{\bar{s}}^{\bar{k}}}{\partial \textbf{u}} \right)^{N_{T}}\left(\textbf{q}, \textbf{u}\right)=0$,  hence  $\boldsymbol{\psi}^{N_T}=0$ because 
 $\textbf{G}$ is invertible.  Then, for each $k<N_T$ Eq.(\ref{discradjpart}) is of the form of
  \begin{linenomath*}$$ \textbf{G}^{\dagger}\boldsymbol{\psi}^{k}={\mathcal F}(\theta,\textbf{q},\,\textbf{u}_{\mathbf{q}},\,\textbf{C},\boldsymbol{\psi}^{k+1},...,\boldsymbol{\psi}^{N_T})$$
  \end{linenomath*}
and has exactly one solution in ${  R}^{N_{Sp}}$ entirely determined by $\boldsymbol{\psi}^{k+1},...,\boldsymbol{\psi}^{N_T}$, $\textbf{q}$, $\theta$  and $\textbf{C}$.
All these $\boldsymbol{\psi}^{k}$ form the unique solution of  Eq.(\ref{Adjequ})  in ${\mathcal D}(A^*)$, which we call $\boldsymbol{\psi}_{\textbf{q}}$.

\section{The derivatives of  $E$ w.r.t  the parameters}
\label{derss}

 The  adjoint problem (\ref{Adjequ})   depends on the differential  of $f$ w.r.t. $\textbf{u}$, and determining  the  gradient of $E$ w.r.t. $\textbf{q}$ 
also needs the derivatives of $f$, $A$, and  $\textbf{r}$   w.r.t. the entries $q_h$ of $\textbf{q}$. Though standard algebra returns these derivatives, two points are worth being mentioned.

 First, for $j=2,3$, the $\partial E/ \partial p_{j}^{(i)}$ are obtained by applying chain rule to (\ref{LE}) and  (\ref{interpb}).

Then, the derivative w.r.t. $\alpha$ involves
$\frac{\partial {\mathcal I}^0}{\partial \alpha}$,
  $\frac{\partial \left(\!{\mathcal I}^{j\!-\!1,k\!-\!1}\!-\!{\mathcal I}^{j\!-\!1,k\!-\!1}\!\right)}{\partial \alpha}$ and $\frac{\partial f_{\bar{s}}^{\bar{k}}}{\partial \alpha}$
 for which we need  the
digamma function $\frac{\Gamma'}{\Gamma}$ \cite{Abr} 
\begin{linenomath*}
\begin{equation}
\frac{\Gamma'}{\Gamma}(z+1)=\gamma
 \sum_{k=1}^{\infty}\frac{z}{k(z+k)},
  \label{gamp}
\end{equation}
\end{linenomath*}
where   $\gamma$ is Euler constant. We obtain
\begin{linenomath*}
\begin{equation}
\frac{ \partial {\mathcal I}^{0}}{\partial \alpha}={\mathcal I}^{0} \left [ -\ln (\Delta t)+\frac{\Gamma'}{\Gamma}(3-\alpha) \right],
 \nonumber
\end{equation}
\end{linenomath*}
and
\begin{linenomath*}
\begin{equation}
\begin{gathered}
\frac{\partial{\mathcal I}^{j, k}}{\partial \alpha}={\mathcal I}^{0}
\left [(j+1)^{2-\alpha}-2j^{2-\alpha}+(j-1)^{2-\alpha} \right]
\left ( -\ln (\Delta t)+ \frac{\Gamma'}{\Gamma}(3-\alpha) \right )\\
- \left [\ln(j+1)(j+1)^{2-\alpha}-2\ln(j)j^{2-\alpha}+
\ln(j-1)(j-1)^{2-\alpha} \right ]\quad \mbox{for}\quad 1<j<k,\\
\frac{\partial{\mathcal I}^{k, k}}{\partial \alpha}={\mathcal I}^{0}
\left [ (2-\alpha)k^{1-\alpha}-k^{2-\alpha}+(k-1)^{2-\alpha} \right]
\left ( -\ln (\Delta t)+ \frac{\Gamma'}{\Gamma}(3-\alpha) \right )\\
- \left [k^{1-\alpha}+(2-\alpha)\ln(k)k^{1-\alpha}-\ln(k)k^{2-\alpha}+
\ln(k-1)(k-1)^{2-\alpha} \right ].
\end{gathered}
\nonumber
\end{equation}
\end{linenomath*}

\end{document}